\documentclass[preprint2, longabstract]{aastex}

\usepackage{multirow}
\usepackage{color}

\shorttitle{Chemical abundances in a $\log N$(H\,{\sc i})~=~22 QSO-DLA} 
\shortauthors{Guimar\~aes et al.}

\begin{document}

\title{Metallicities, dust and molecular content of a QSO-Damped Lyman-$\alpha$ system reaching 
$\log N$(H\,{\sc i})~=~22: \\ An analog to GRB-DLAs}

\author{R. Guimar\~aes}
\affil{Programa de Modelagem Computacional - SENAI - Cimatec, 41650-010 Salvador, Bahia, Brasil}
\email{rguimara@eso.org}
\author{P. Noterdaeme and P.  Petitjean}
\affil{UPMC-CNRS, UMR7095, Institut d'Astrophysique de Paris, 98bis bd Arago, F-75014 Paris, France}
\author{C. Ledoux}
\affil{European Southern Observatory, Alonso de C\'ordova 3107, Casilla 19001, Vitacura, Santiago 19, Chile}
\author{R. Srianand}
\affil{Inter-University Centre for Astronomy and Astrophysics, Post Bag 4, Ganeshkhind, Pune 411 007, India}
\author{S. L\'opez}
\affil{Departamento de Astronom\'ia, Universidad de Chile, Casilla 36-D, Santiago, Chile}
\author{H. Rahmani}
\affil{Inter-University Centre for Astronomy and Astrophysics, Post Bag 4, Ganeshkhind, Pune 411 007, India}

\begin{abstract}
We present the elemental abundance and H$_2$ content measurements of a Damped Lyman-$\alpha$ (DLA) system with an extremely large H\,{\sc i} column density, $\log N$(H\,{\sc i})~(cm$^{-2}$)~=~22.0$\pm$0.10, at $z_{\rm abs} = 3.287$ towards the QSO SDSS J\,081634$+$144612. We measure column densities of H$_2$, C\,{\sc i}, C\,{\sc i}$^{\star}$, Zn\,{\sc ii}, Fe\,{\sc ii}, Cr\,{\sc ii}, Ni\,{\sc ii} and Si\,{\sc ii} from a high signal-to-noise and high spectral resolution VLT-UVES spectrum. The overall metallicity of the system is [Zn/H]~=~$-1.10\pm0.10$ relative to solar. Two molecular hydrogen absorption components are seen at $z=3.28667$ and 3.28742 (a velocity separation of $ \approx 52$~km s$^{-1}$) in rotational levels up to $J=3$. We derive a total H$_2$ column density of log~$N$(H$_2$)~(cm$^{-2}$)~=~18.66  and a mean molecular fraction of $f$~=~$2N($H$_2)/[2N$(H$_2)+N($H\,{\sc i}$)] = 10^{-3.04\pm0.37}$, typical of known H$_2$-bearing DLA systems.  From the observed abundance ratios we conclude that dust is present in the Interstellar Medium (ISM) of this galaxy, with a enhanced abundance in the H$_2$-bearing clouds. However, the total amount of dust along the line of sight is not large and does not produce any significant reddening of the background QSO.
The physical conditions in the  H$_2$-bearing clouds are constrained directly from the column densities of H$_2$ in different rotational levels, C\,{\sc i} and C\,{\sc i}$^{\star}$. The kinetic temperature is found to be $T \approx 75$~K and the particle density lies in the range $n_{\rm H}$~=~50$-$80~cm$^{-3}$. 
The neutral hydrogen column density of this DLA is similar to the mean H\,{\sc i} column density of DLAs observed at the redshift of $\gamma$-ray bursts (GRBs). We explore the relationship between GRB-DLAs and high column density end of QSO-DLAs finding that the properties (metallicity and depletion) of DLAs with $\log N$(H\,{\sc i})~$>$~21.5 in the two populations do not appear to be significantly different.
\end{abstract}


\section{Introduction}

Despite accounting for only a small fraction of all the baryons in the Universe (see, e.g., 
Petitjean et al. 1993), the physical state of the neutral and molecular phases of the interstellar 
medium is a crucial ingredient of galaxy formation. These gaseous phases are at any redshift the 
reservoir of gas available for star formation. At high redshift, most of the neutral hydrogen mass 
is revealed by the damped Lyman-$\alpha$ absorption system (DLAs) detected in the spectra of 
background quasars (see e.g. Wolfe et al. 2005 for a review). Since DLAs are easy to identify 
in QSO spectra and the H\,{\sc i} column densities can be measured accurately, it is possible to 
derive the cosmological mass density of the neutral gas at different redshifts, independent of 
the exact nature of the absorbers, provided a sufficiently large number of background quasars are observed (see Prochaska \& Wolfe 2009, Guimar\~aes et al. 2009, Noterdaeme et al. 2009b).

Key results of DLA surveys include the indication that the $N$(H\,{\sc i}) frequency distribution deviates significantly from a single power-law with $f$($N$(H\,{\sc i})) sharply steepening at $\log N$(H\,{\sc i})~$>$~21.5. However, Zwaan \& Prochaska (2006) used CO emission maps in the nearby Universe to show that the H$_2$ column density distribution function is a continuous extension of the H\,{\sc i} distribution for high column densities. 
The transition happens at log~$N$(H\,{\sc i})~$\sim$~22 which is the approximate column density associated with the conversion from H\,{\sc i} into H$_2$ (e.g., Schaye 2001).

The slope of the column density distribution at large $N$(H\,{\sc i}), $\alpha \sim −3.5$, implies that systems with very large column density are very rare. Indeed, until very recently\footnote{In the last stages of this work, the discovery of another DLA with log~$N$(H~{\sc i})~$\ge$~22 was reported by Noterdaeme et al. (2012) and Kulkarni et al. (2012).},  only one DLA system with $\log N$(H\,{\sc i})~$\sim$~22 was reported in the literature: the system at $z_{\rm abs}=3.287$ towards SDSS J\,081634$+$144612 (from the SDSS-II DLA catalog, Noterdaeme et al. 2009b).
However, DLAs with such high column densities are frequently detected at the redshift of $\gamma$-ray bursts (e.g. Savaglio et al. 2003; Vreeswijk et al. 2004; Fynbo et al. 2006; Jakobsson et al. 2006; Prochaska et al. 2007; Fynbo et al. 2009; Ledoux et al. 2009; Savaglio 2010). 
Therefore, comparing the chemical and physical properties of the strongest QSO-DLAs to that of GRB-DLAs may provide clues to understand the nature of these absorbers. 

We present here a detailed study of the gas-phase abundances of metals, dust and molecules in the QSO-DLA towards SDSS J\,081634$+$144612, based on high spectral resolution data.
The paper is organized as follows. In Sect.~2 we provide details of observations and data reduction.
The hydrogen column density of the DLA, the metal, dust and molecular content are discussed in Sect.~3. The physical state of the H$_2$ bearing gas component is discussed in Sect.~4. In Sect.~5, we finally investigate the connection between QSO-DLAs and GRB-DLAs.

\section{Observations}
We observed quasar SDSS J\,081634$+$144612 twice, in September 2008 and April 2009, with the high-resolution Ultraviolet and Visual Echelle spectrograph (UVES, Ballester et al. 2000) mounted on the ESO Kueyen VLT-UT2 8.2~m telescope at Cerro Paranal, Chile. Observations have been performed under programs 081.A-0334(A), PI. S. L\'opez in visitor mode, and 282.A-5030(A), PI. P. Noterdaeme in service mode. Ten exposures were taken for a total of 12.4~hours exposure time: nine exposures using Dichroic 2 with a setting 437+760~nm plus one 5400\,s exposure with the red arm centered at 550~nm that covers the Lyman-$\alpha$ absorption.
A slit width of 1~arcsec and 2x2 pixel binning were used, resulting in a spectral resolution of 50,000. 

The quasar spectrum was reduced using the UVES pipeline (see e.g. Ledoux et al. 2003 for details). The main characteristics of the pipeline are to perform a precise inter-order background subtraction, especially for master flat-fields, and to allow for an optimal extraction of the object signal 
rejecting cosmic rays and performing sky subtraction at the same time. The pipeline products were checked step by step. 
The wavelength scale of each reduced spectrum was then converted to vacuum-heliocentric values and the spectra rebinned to a constant wavelength step. 
No further rebinning was performed during the analysis of the whole spectrum. Individual 1-D exposures were scaled, weighted and combined together.

In order to derive the physical parameters of the absorption features, we fit the metal absorption profiles with multiple Voigt profiles using VPFIT (Carswell et al. 1987). The continuum level was obtained locally in the vicinity of each metal absorption feature. Molecular hydrogen features were fitted altogether using fit/lyman and after normalizing the corresponding region of the Lyman-$\alpha$ forest.
Atomic data for metal species and H$_2$ are, respectively, from Morton (2003) and Bailly et al. (2010). 
In the following, solar abundances are taken from Lodders (2003). The origin of the velocity scale ($v$~=~0~km~s$^{-1}$) is set at the redshift of the single C\,{\sc i}  component, $z=3.28746$.

\section{Abundances}

\subsection{H\,{\sc i} and metal content}

\begin{figure}
\includegraphics[scale=1]{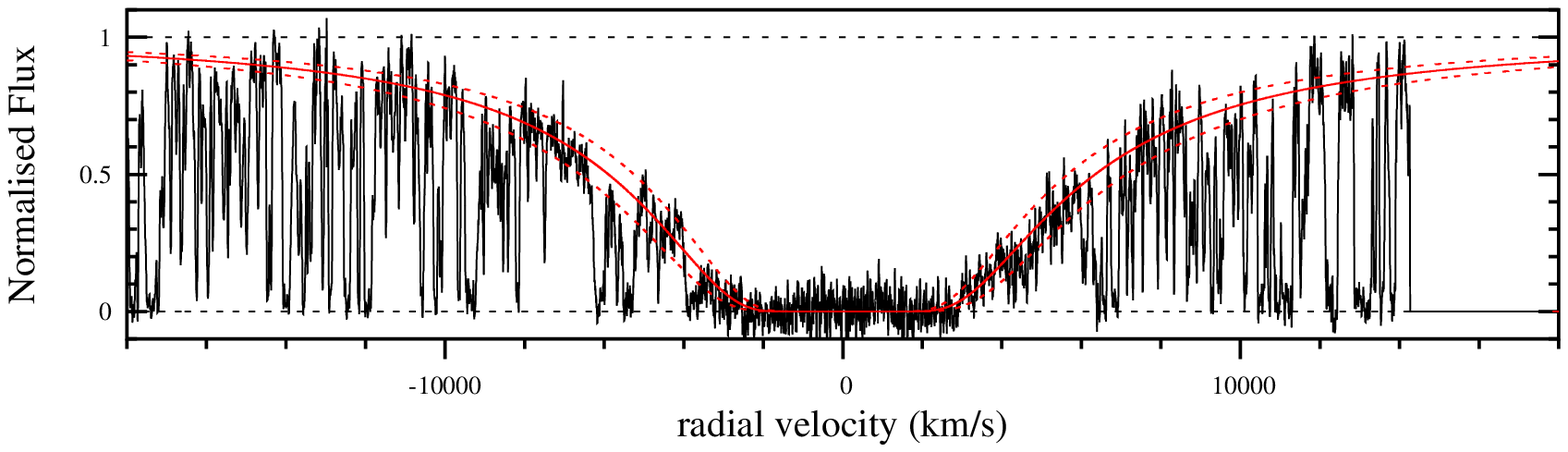}
\caption{H\,{\sc i} Lyman-$\alpha$ profile of the DLA at $z = 3.286$ toward QSO SDSS J\,081634$+$144612. The overplotted solid line and accompanying dashed lines correspond to the best fit solution log~$N$(H\,{\sc i})~=~22.0$\pm$0.10. The origin of the velocity scale is taken at $z = 3.287$.}
\label{dla}
\end{figure}

The neutral hydrogen column density was measured from the fit of the damping  wings of the Lyman-$\alpha$ absorption at $z = 3.287$. We find $\log N$(H\,{\sc i}) (cm$^{-2}$)~=~22.0$\pm$0.10.
The observed Damped Lyman-$\alpha$ absorption together with the best fitted Voigt profile is shown in Figure~\ref{dla}. The dashed lines indicate the profiles corresponding to $\log N$(H\,{\sc i})~=~21.9 and 22.1. 

\begin{figure}
\renewcommand{\tabcolsep}{1pt}
\centering
\begin{tabular}{ccc}
\includegraphics[bb=219 357 393 620, clip=, angle=90, width=0.25\hsize]{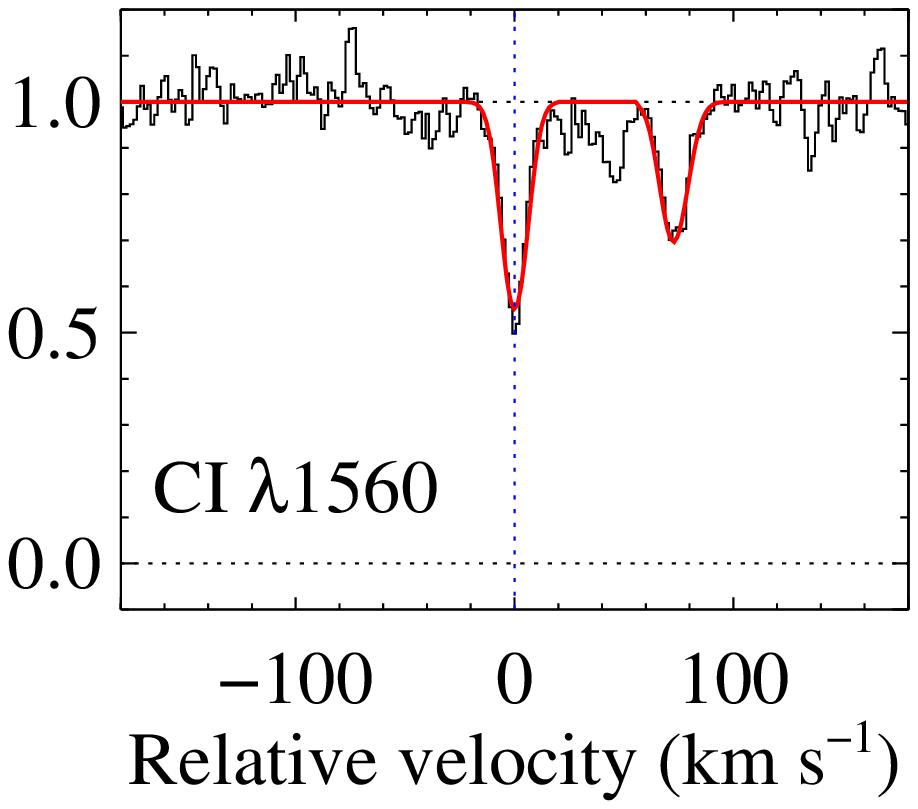} &
\includegraphics[bb=219 357 393 620, clip=, angle=90, width=0.25\hsize]{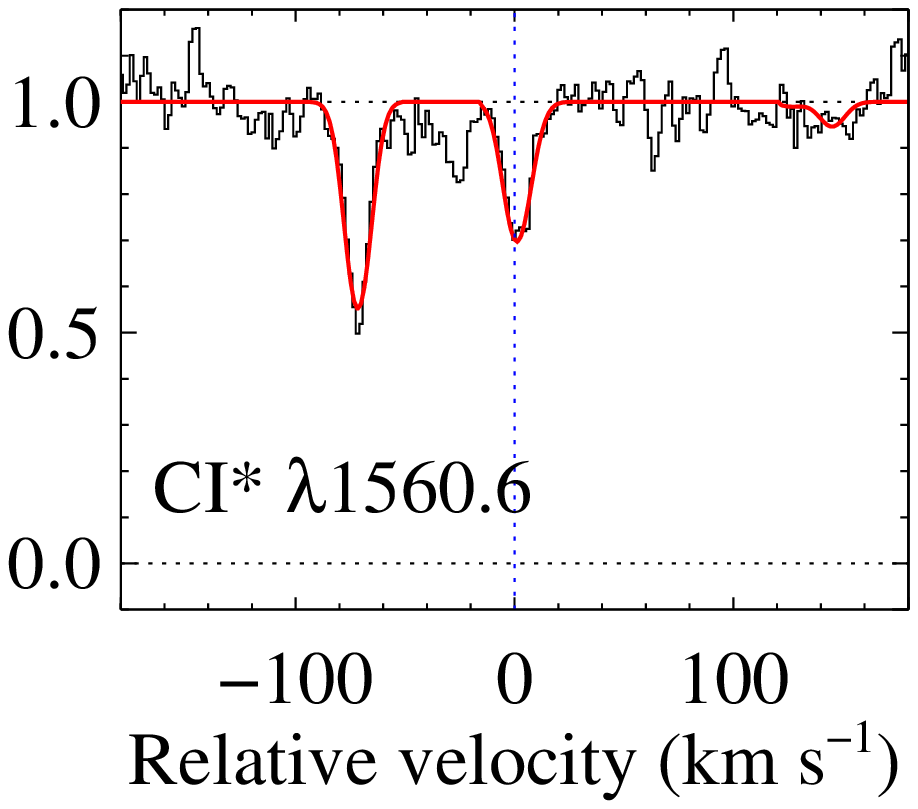} &
\includegraphics[bb=219 357 393 620, clip=, angle=90, width=0.25\hsize]{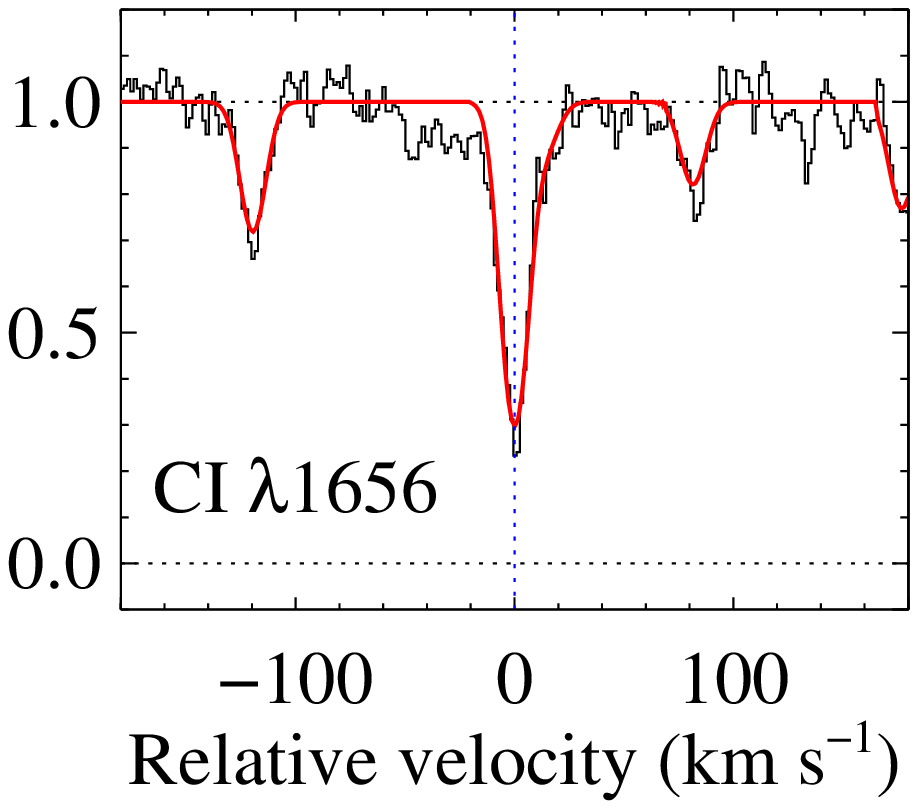} \\
\includegraphics[bb=219 357 393 620, clip=, angle=90, width=0.25\hsize]{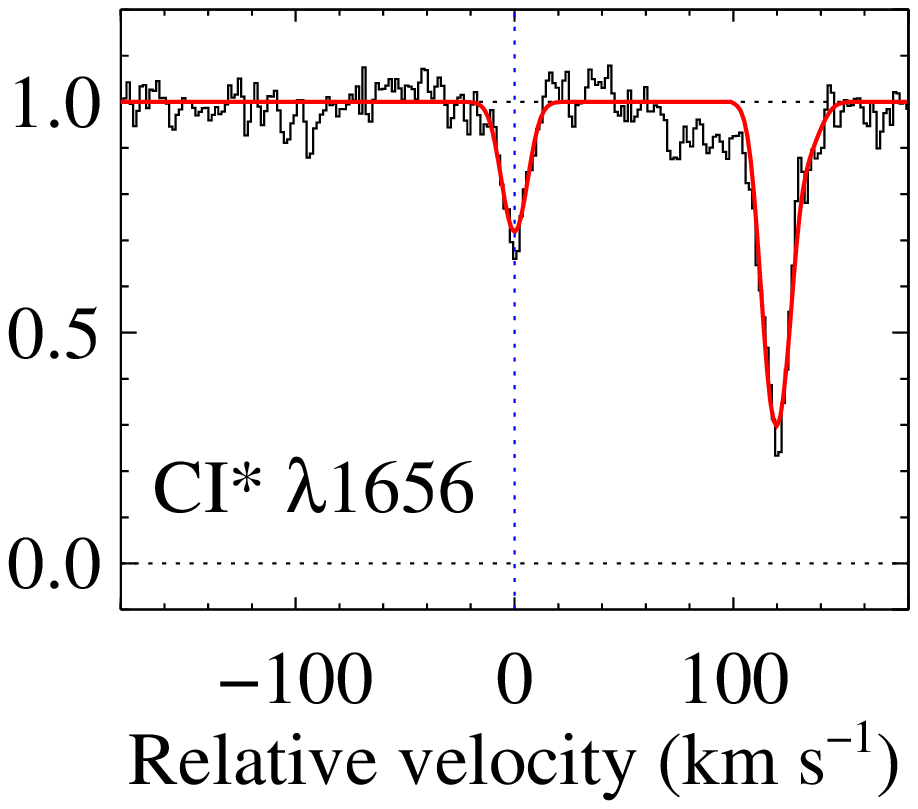} &
\includegraphics[bb=219 357 393 620, clip=, angle=90, width=0.25\hsize]{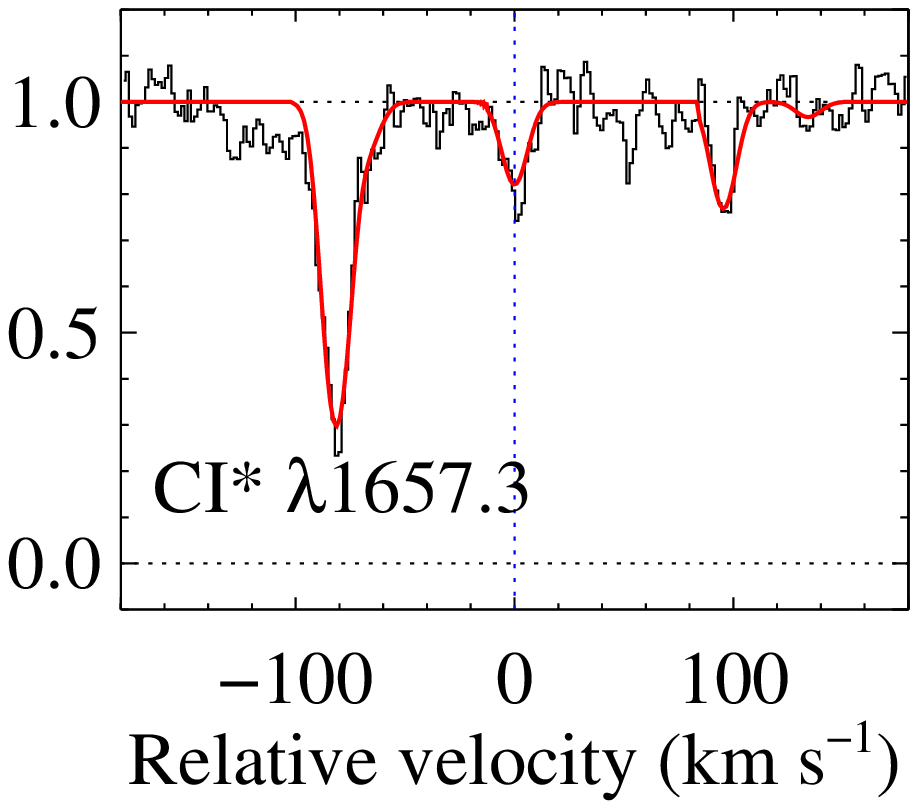} &
\includegraphics[bb=219 357 393 620, clip=, angle=90, width=0.25\hsize]{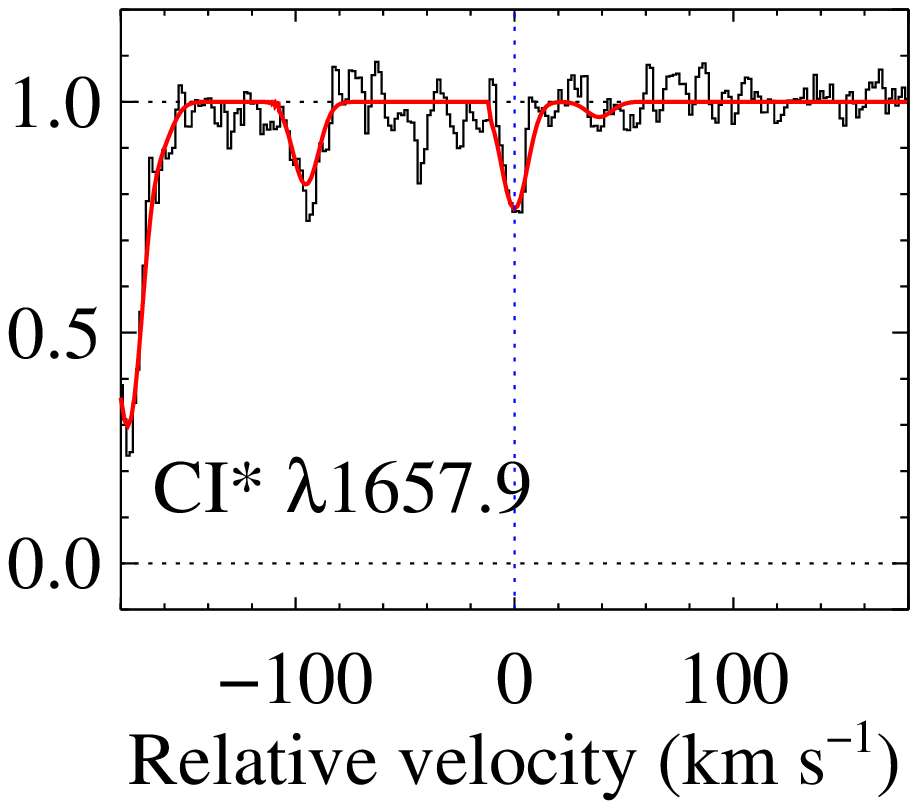} \\
\includegraphics[bb=219 357 393 620, clip=, angle=90, width=0.25\hsize]{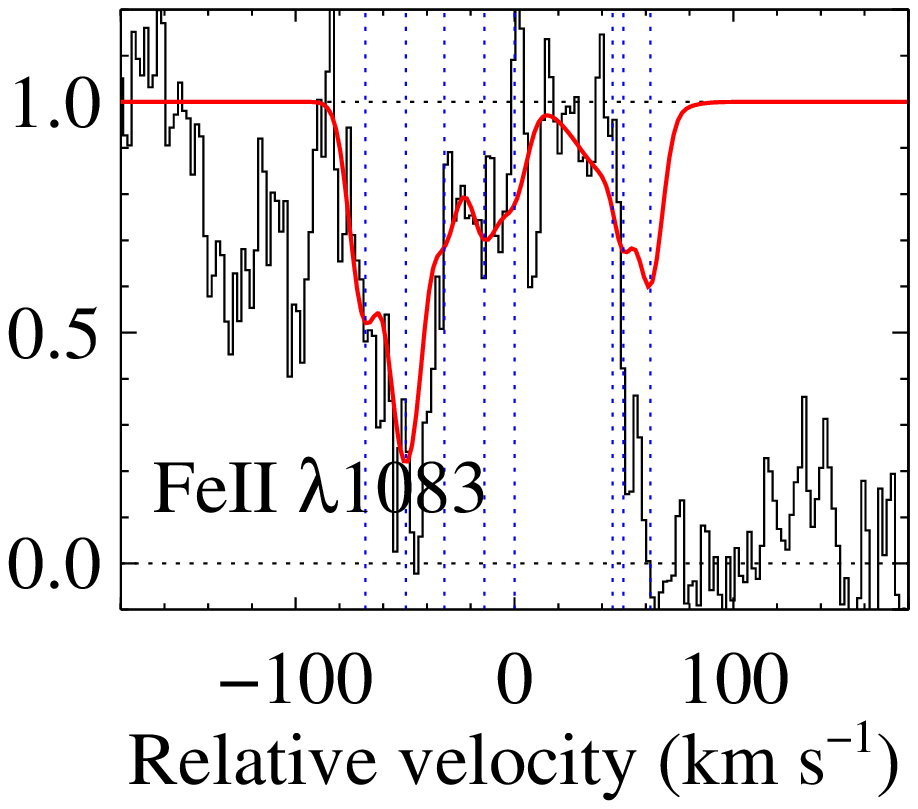} &
\includegraphics[bb=219 357 393 620, clip=, angle=90, width=0.25\hsize]{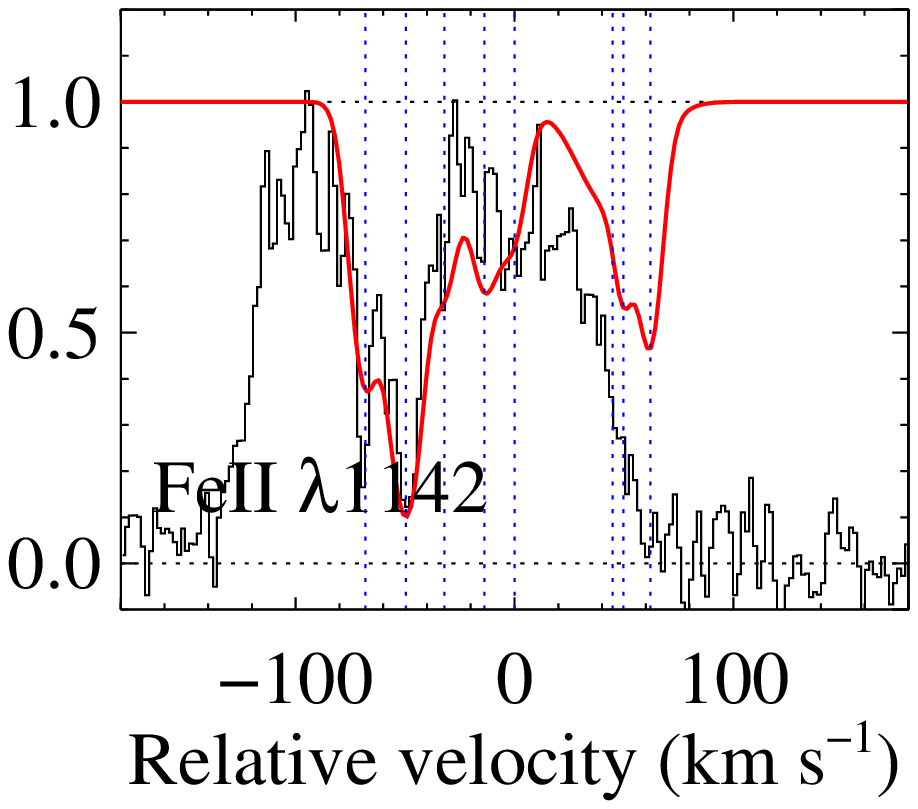} &
\includegraphics[bb=219 357 393 620, clip=, angle=90, width=0.25\hsize]{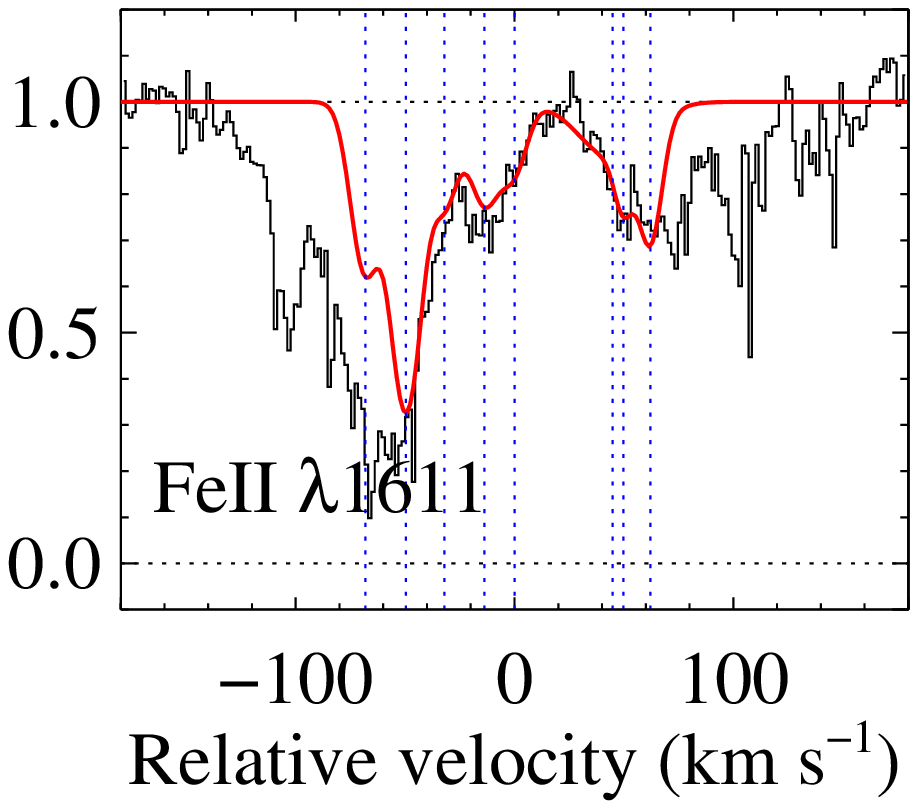} \\
\includegraphics[bb=219 357 393 620, clip=, angle=90, width=0.25\hsize]{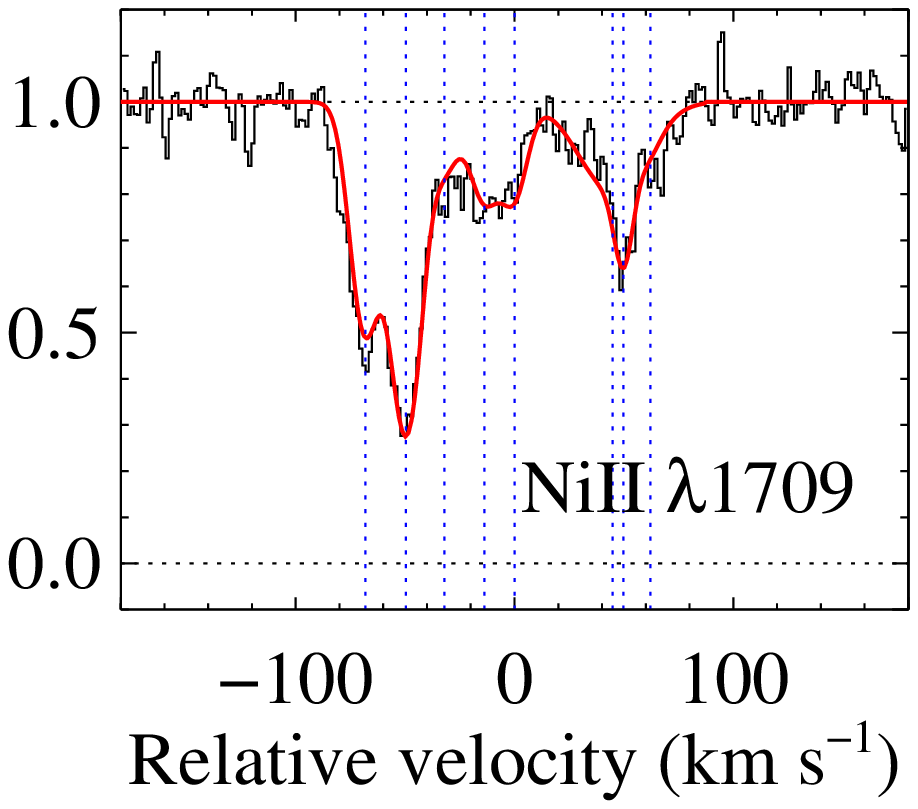} &
\includegraphics[bb=219 357 393 620, clip=, angle=90, width=0.25\hsize]{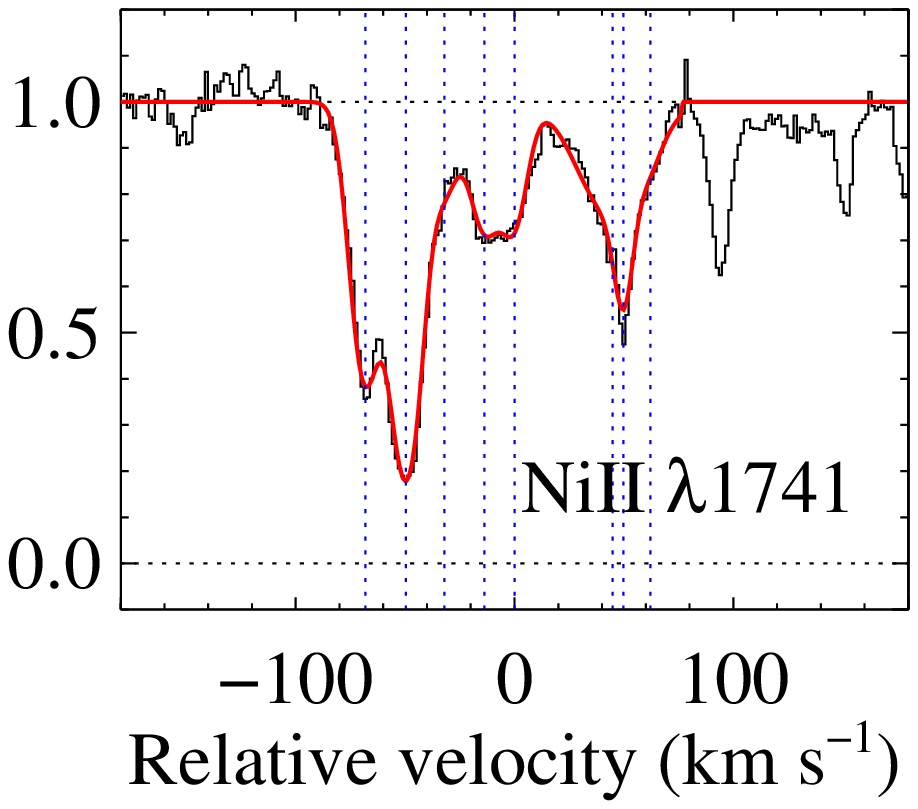} &
\includegraphics[bb=219 357 393 620, clip=, angle=90, width=0.25\hsize]{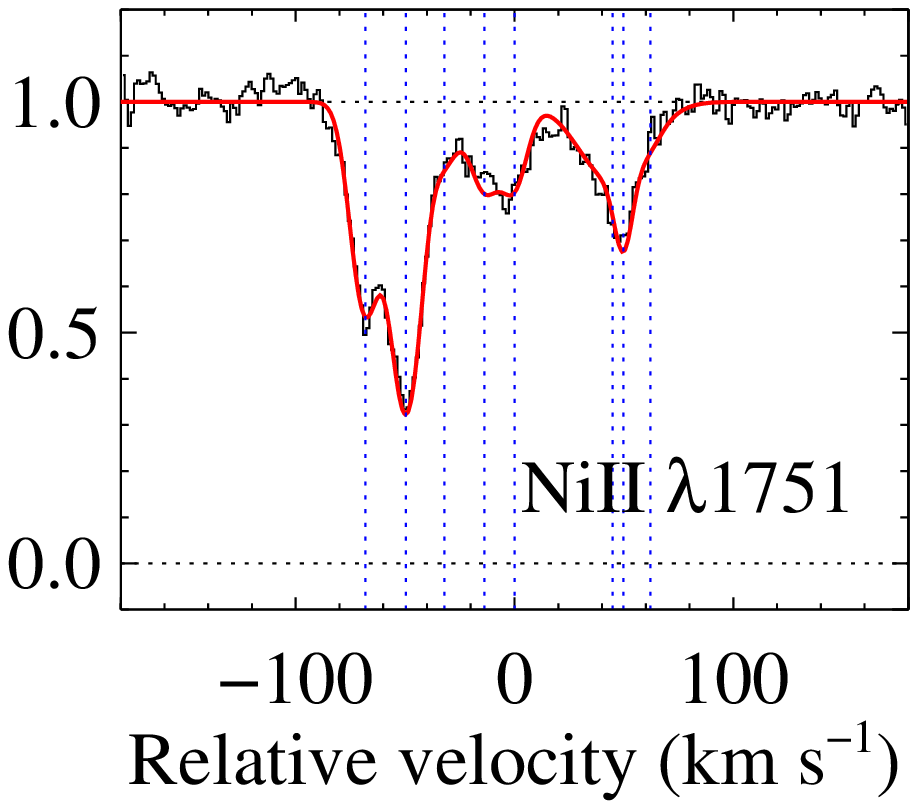} \\
\includegraphics[bb=219 357 393 620, clip=, angle=90, width=0.25\hsize]{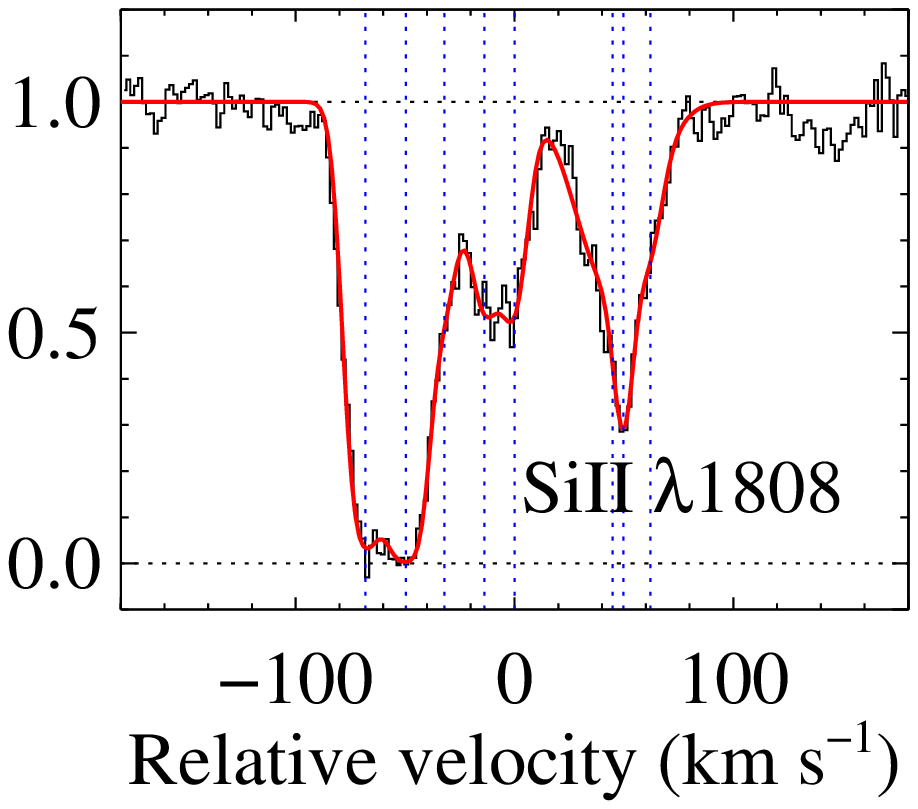} &
\includegraphics[bb=219 357 393 620, clip=, angle=90, width=0.25\hsize]{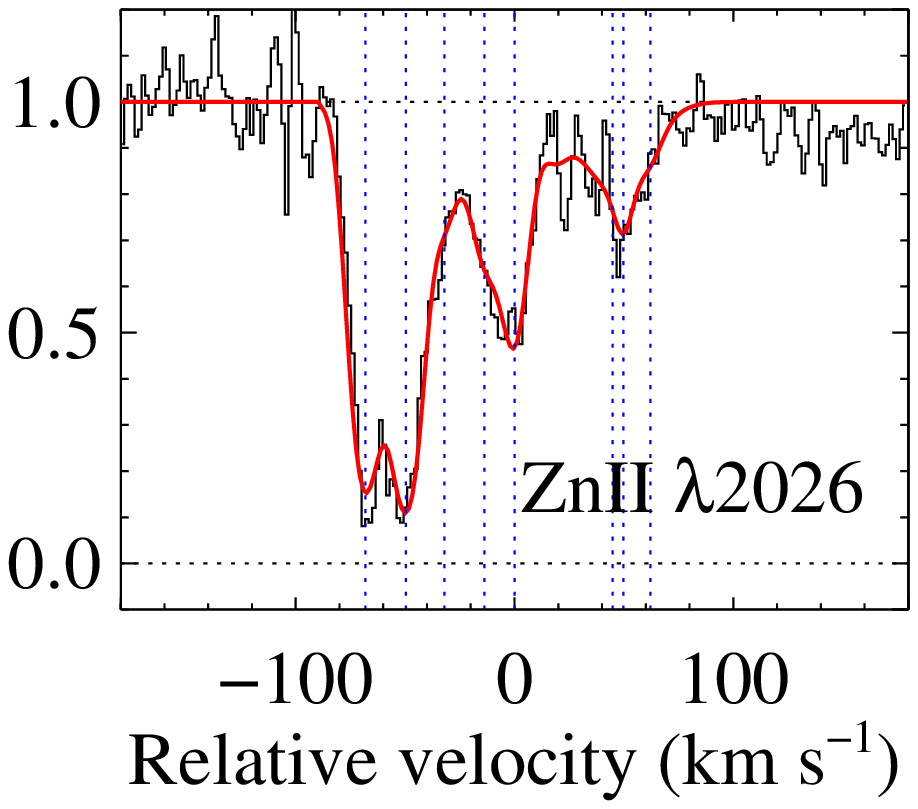} &
\includegraphics[bb=219 357 393 620, clip=, angle=90, width=0.25\hsize]{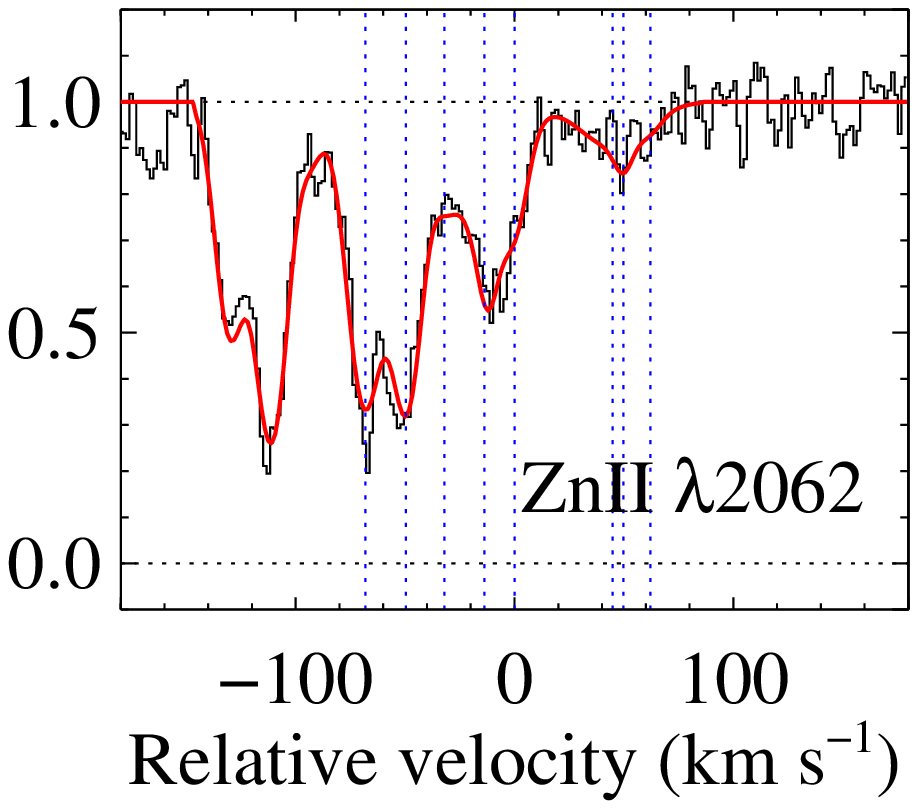} \\
\includegraphics[bb=165 357 393 620, clip=, angle=90, width=0.25\hsize]{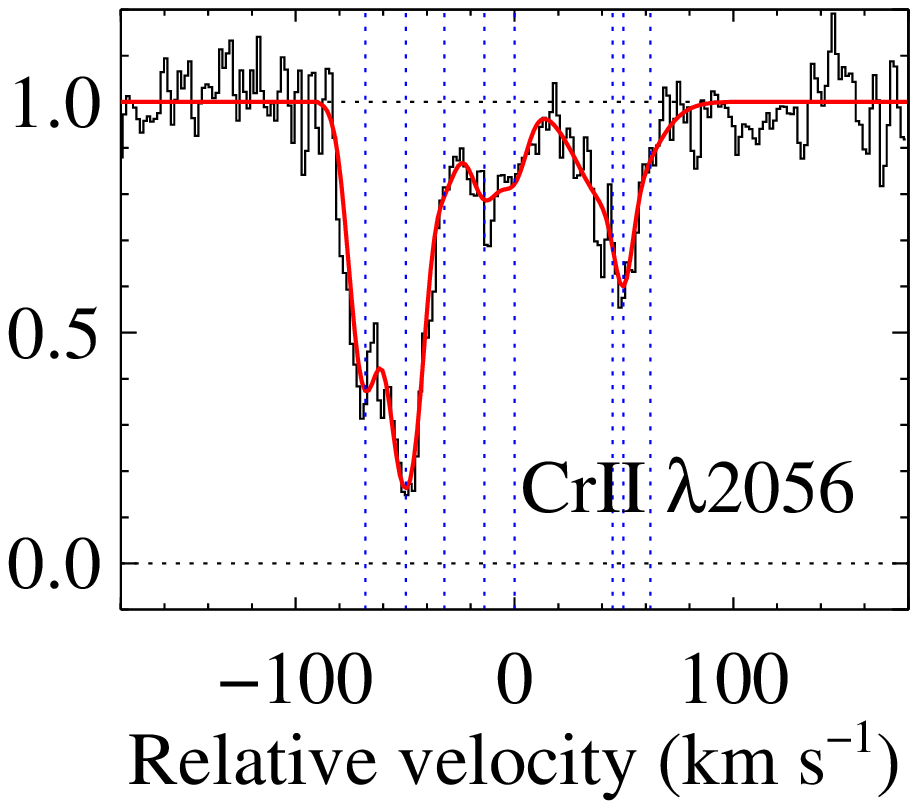} &
\includegraphics[bb=165 357 393 620, clip=, angle=90, width=0.25\hsize]{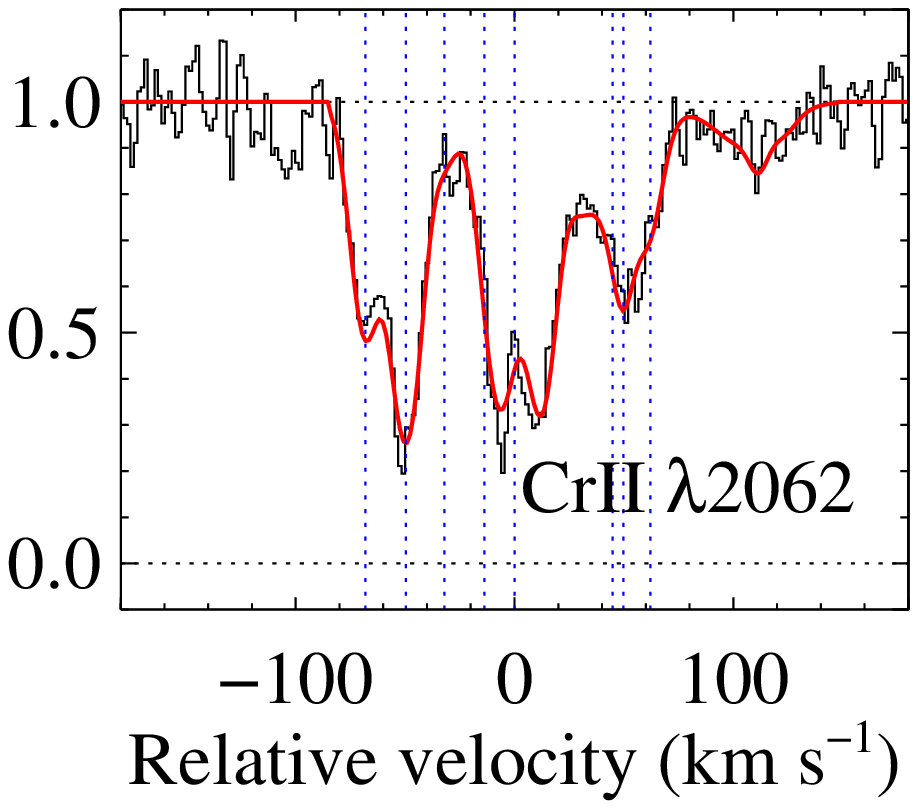} &
\includegraphics[bb=165 357 393 620, clip=, angle=90, width=0.25\hsize]{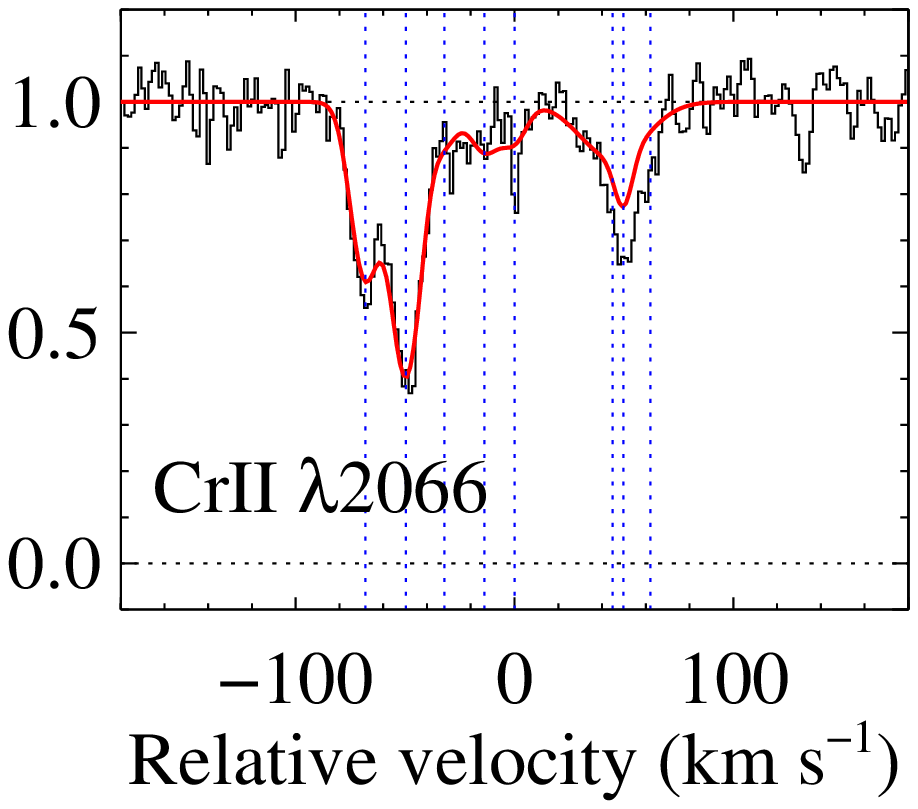} \\
\end{tabular}
\renewcommand{\tabcolsep}{6pt}
\caption{Observed absorption profiles of C\,{\sc i}, C\,{\sc i}*, Zn\,{\sc ii}, Fe\,{\sc ii}, Cr\,{\sc ii}, Ni\,{\sc ii}, Si\,{\sc ii} together with fitted profiles. The dashed vertical lines indicate the positions of the velocity components. The origin of the velocity scale is taken at $z = 3.287456$, the redshift of the C\, {\sc i} component. Regions affected by blends (see Fe\,{\sc ii} lines as well as the reddest clump of Cr\,{\sc ii}$\lambda$2066) are not considered in the fitting process. Additional absorption seen at +150~km\,s$^{-1}$ the C\,{\sc i}$^\star$ $\lambda$1560.6 panel and in the red wing of C\,{\sc i}$\lambda$1656 and C\,{\sc i}$^\star$ $\lambda$1657.9 are due to C\,{\sc i}$^{\star\star}$.}
\label{ajuste}
\end{figure}

We detect absorption lines of C\,{\sc i}, C\,{\sc i}$^\star$, Zn\,{\sc ii}, Fe\,{\sc ii}, Cr\,{\sc ii}, Ni\,{\sc ii} and Si\,{\sc ii}, spread over about 150~km\,s$^{-1}$.  As can be seen in Fig.~\ref{ajuste}, the absorption profiles are not strongly saturated except may be for Si\,{\sc ii}$\lambda$1808. We are therefore confident that our column density determinations are robust. The fit to the absorption lines are overplotted on the Figure as red solid lines. The singly ionized species are expected to be the dominant contributors to the abundances of the corresponding elements in H\,{\sc i} clouds with such high column densities.

Voigt profile fitting is performed simultaneously for all the absorption lines keeping the same number of components having same redshifts and Doppler parameters for all singly ionized species. 
Note that Zn\,{\sc ii}$\lambda$2026 is blended with Mg\,{\sc i}$\lambda$2026. The contributions of the latter is taken into account in the fits and found to have negligible influence on the derived N(Zn\,{\sc ii}).

Eight velocity components are necessary to model the profiles (see Fig.~\ref{ajuste}). 
We report in Table \ref{parameters} the results of the fits, column density and Doppler parameter,
for each of the components. Because of saturation effects, the Si\,{\sc ii} column density should be considered a lower limit. However, given the shape of the absorption, we are confident that the true value cannot be much larger.  The column density of singly ionized iron, despite their blending, could be reliably measured.

Although the profile decomposition may not be unique, three distinct clumps can be identified (\#1, 2 and 3 from blue to red) at mean redshifts of 3.28661, 3.28735 and 3.28814, and made of respectively, 3, 2 and 3 components.
H$_2$ absorption is detected in \#1, which is the strongest clump in metal species, but most of the H$_2$ is found associated to the C\,{\sc i} component in clump \#2.
Unfortunately, we cannot determine $N$(H\,{\sc i}) in each clump. The mean metallicity\footnote{Metallicities 
are given relative to solar: ${\rm [X/H]} = {\rm log}(N_{\rm X}/N_{\rm HI})_{\rm DLA} - {\rm log}({\rm X/H})_{\odot}$} in the cloud, derived by adding the column densities of the eight components are: [Zn/H]~=~$-1.10\pm0.10$, [Si/H]~$\geq$~$-1.23\pm0.10$, [Cr/H]~=~$-1.58\pm0.10$, [Ni/H]~=~$-1.68\pm0.10$ and [Fe/H]~=~$-1.58\pm0.10$.

\subsection{Dust}

\begin{figure}
\centering
\includegraphics[width=8.5cm, height=8.5cm,bb=70 200 480 570]{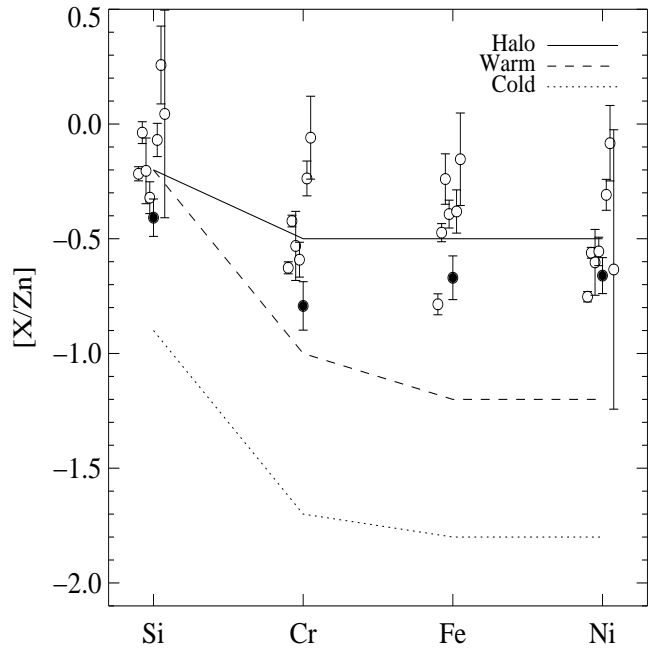} 
\caption{Depletion pattern of elements relative to zinc measured in individual components in the 
DLA toward SDSS J\,081634$+$144612. The points are slightly shifted along the x-axis according to their position in the system (for clarity purpose only). 
The filled circle indicates the component associated with the strongest H$_2$ absorption.
The dotted, dashed and continuous lines represent the typical relative abundances observed in, respectively, cold or warm gas in the Galactic disk and diffuse gas in the Galactic Halo, from Welty et al. (1999). 
}
\label{XZn}
\end{figure}

In the ISM of the Galaxy, zinc is virtually undepleted onto dust grains when Si, Cr and Ni are.
We find in the present DLA mean relative abundances:  [Si/Zn]~=~$-0.13\pm0.03$, 
[Cr/Zn]~=~$-0.48\pm0.02$, [Ni/Zn]~=~$-0.58\pm0.02$ and  [Fe/Zn]~=~$-0.48\pm0.02$ indicating that the overall depletion of Si, Cr, Ni and Fe is similar to what is seen in the gas from the halo of our Galaxy (Welty et al. 1999). This is also typical of DLAs where H$_2$ is detected (Noterdaeme et al. 2008). 
From Fig.~\ref{XZn}, which presents the depletion patterns relative to zinc observed component by component, it is clear however that the depletion is enhanced in the main H$_2$-bearing component.  This situation is similar to that of the H$_2$-bearing component towards Q~0013$-$004 (Petitjean et al. 2002), where higher depletion factors are seen in the H$_2$ components.

\begin{figure}
\centering
\includegraphics[angle=90,width=\hsize]{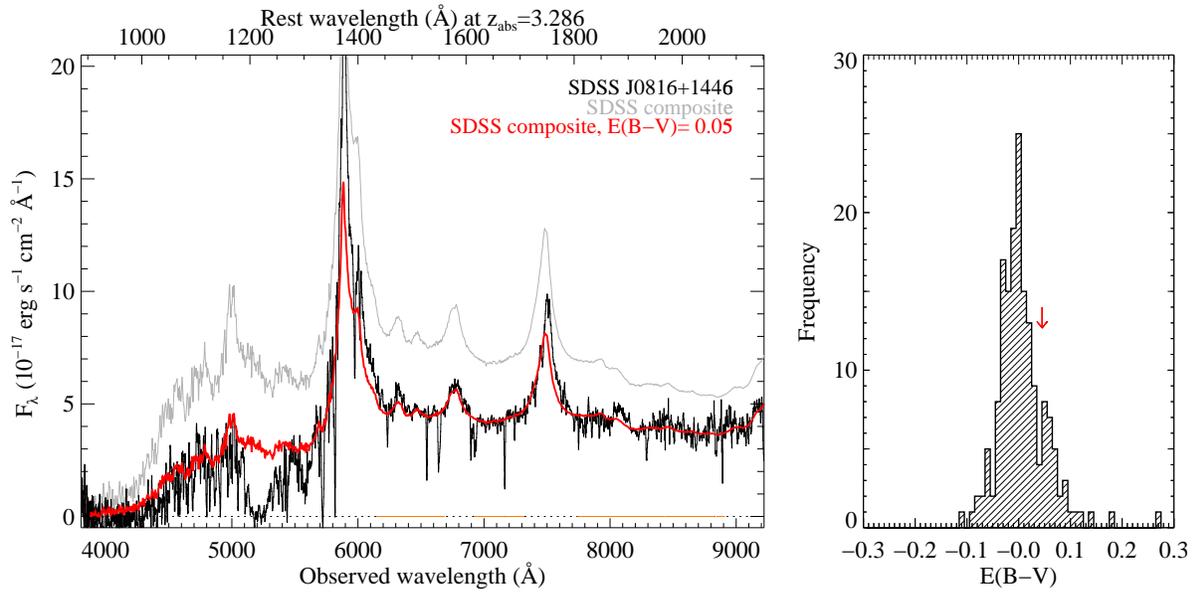} 
\caption{ {\sl Left}: The SDSS spectrum of J\,081634$+$144612 (black) is shown together with the unreddened SDSS composite (grey) and the same reddened using a SMC extinction law and E(B-V)~=~0.05 (red). {\sl Right}: Distribution of E(B-V) measured for a control sample of 163 QSOs within $\Delta z = \pm 0.02$ from SDSS J\,081634$+$144612. The arrow indicate the position of the latter.}
\label{ebv}
\end{figure}

From the flux-calibrated SDSS spectrum, it is also possible to estimate the reddening induced by the presence of dust in the DLA to the background QSO light, following the method described in Noterdaeme et al. (2009b). In Fig.~\ref{ebv}, we show that the SDSS spectrum of J\,081634$+$144612 is well matched with the SDSS composite spectrum from Vanden Berk et al. (2001), shifted to the same emission redshift and reddened using a SMC extinction law (Gordon et al. 2003) at $z=3.286$ with \mbox{E(B-V)}~=~0.05$\pm$0.06. The associated uncertainty is obtained from the dispersion measured 
for a control sample of 163 SDSS QSOs from Schneider et al. (2010) with emission redshift within $\pm 0.02$ to that of J\,081634$+$144612. This means that there is no significant reddening of the quasar J\,081634$+$144612. Overall, the measured extinction-to-dust ratio, A$_{\rm v}/N$(H\,{\sc i})~$<5\times10^{-23}$~mag\,cm$^2$ (2\,$\sigma$) is typical of that of the general DLA population (Vladilo et al. 2008). This, together with the presence of H$_2$ in a component with 
higher depletion factor suggests that appreciable fraction of H\,{\sc i} may be associated with components that do not have H$_2$.

\subsection{H$_2$ content}

\begin{figure}
\includegraphics[width=18cm,height=20cm, angle=180]{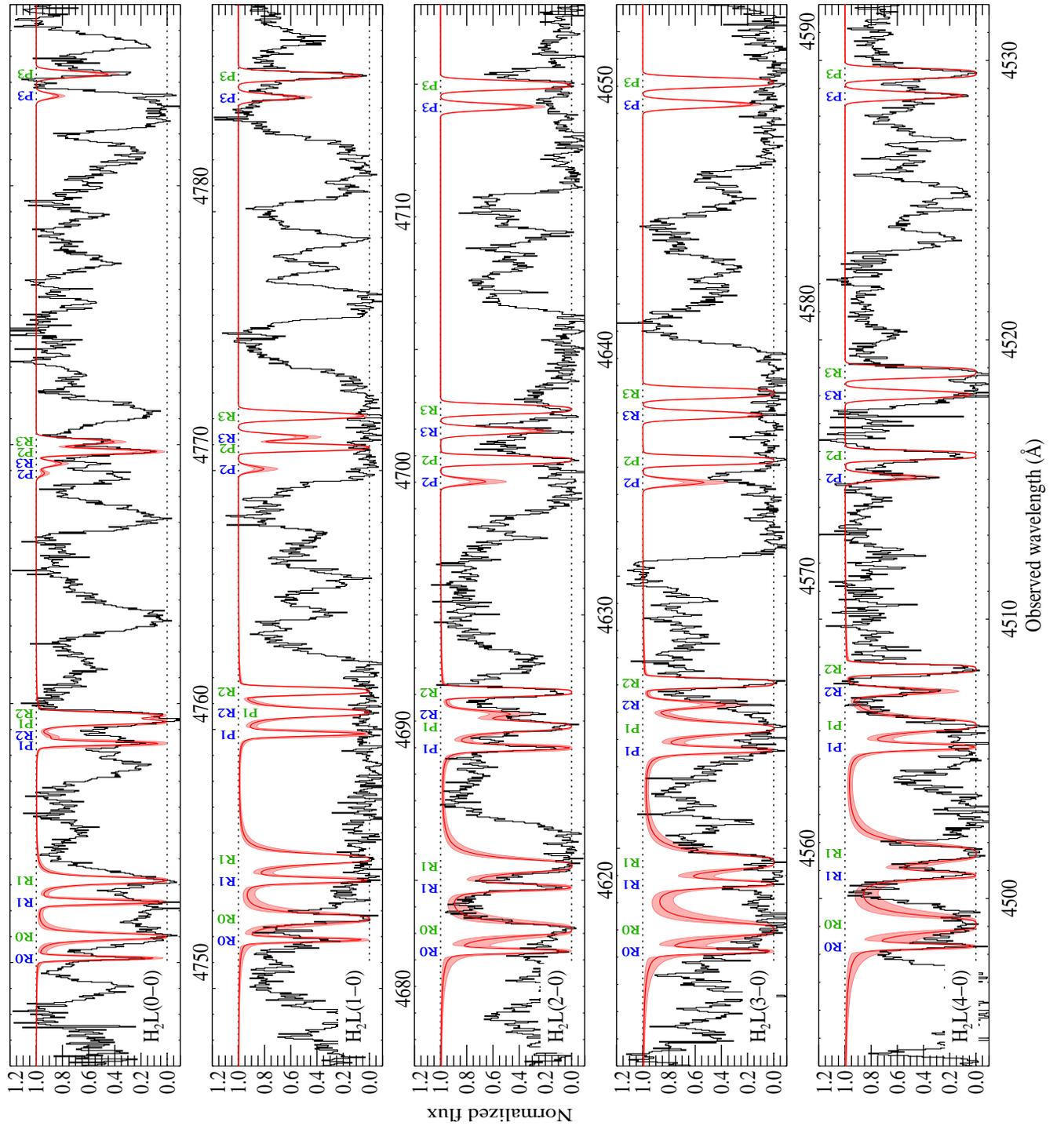} 
\caption{Portions of the SDSS J\,081634$+$144612 normalized spectrum. The best fit to the H$_2$ absorption lines in several Lyman bands (0-0 to 4-0) is superimposed onto the spectrum, with the corresponding uncertainty represented by the shaded area. The fit parameters are given in Table \ref{molecular}. The blue and green labels indicate the branch and the rotational level for the bluest and reddest component of H$_2$, respectively.}
\label{ajuste_H2}
\end{figure}

Molecular hydrogen is detected in two distinct sub-systems at $z_{\rm abs}$~=~3.28667 and 3.28742, 
separated by $\sim$52~km~s$^{-1}$ with absorption lines from rotational levels up to J~=~3 
(see Figure~\ref{ajuste_H2}). The results of the fits to the numerous absorption lines are given 
in Table~\ref{molecular}. The $z_{\rm abs}$~=~3.28742 component alone contains about 90\% of the total 
H$_2$ column density in the absorber, log~$N$(H$_2$)~=~18.62, and coincides with the C\,{\sc i} component. 
This is expected because the energy of the photons that ionize C\,{\sc i} is close to that of photons that dissociate H$_2$.
The total H$_2$ column density integrated over the two components and all rotational levels is 
log~$N$(H$_2$)~=~18.66~$\pm$~0.27~(cm$^{-2}$) corresponding to a molecular fraction of 
log~$f$~=~log~2$\times$ $N$(H$_2$)/(2$\times$ $N$(H$_2$)~+~$N$(H\,{\sc i}))~=~$-3.04 \pm 0.37$ if 
we assume that the totality of neutral hydrogen is associated with the H$_2$ components. 
This value is amongst the lowest observed in H$_2$ bearing DLAs with metallicities [Zn/H]~$>$~$-1.3$ 
(Petitjean et al. 2006, Noterdaeme et al. 2008). This again may indicate that appreciable fraction of H\,{\sc i} may be associated with components that do not have H$_2$ (see also Noterdaeme et al. 2010, Srianand et al. 2010, 2012).

\section{Physical state of the gas}

\subsection{Excitation of H$_2$}

\begin{figure}
\centering
\includegraphics[width=8.5cm, bb=85 175 490 610,clip=]{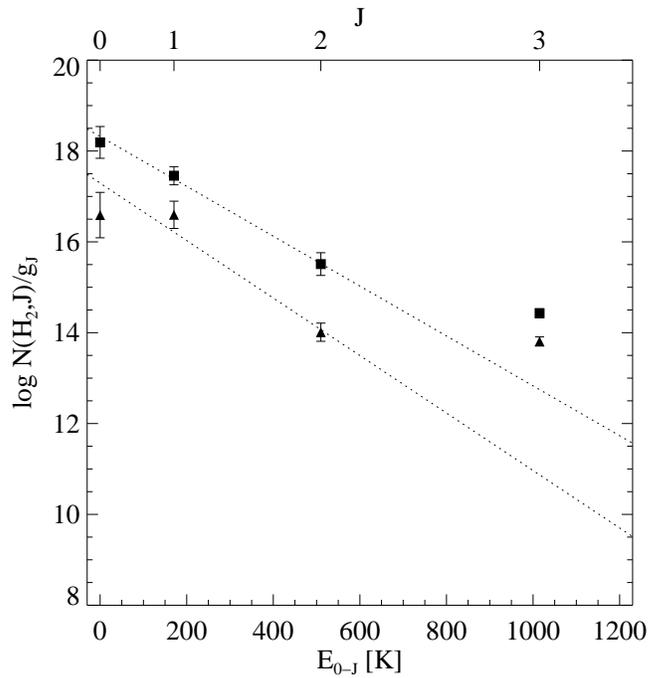} 
\caption{H$_2$ excitation diagram for the 2 components observed toward SDSS J\,081634$+$144612. Triangles correspond to $z_{\rm abs}$ = 3.28667 and squares to $z_{\rm abs}$ = 3.28742. The column density $N_{\rm J}$ divided by the statistical weight, $g_{\rm J}$, is plotted for the J~=~0 up to J~=~3 H$_2$ rotational levels on a logarithmic scale against the excitation energy, $E_{\rm J}$, in K.
The J=0 to J=2 points have been fitted with straight lines using Eq.~\ref{excitation} giving the excitation temperatures $T_{\rm ex}^{012}$ indicated in Table~\ref{molecular}.}
\label{excitation_H2}
\end{figure}

From the detection of H$_2$ in different rotational levels (J=0 to J=3, see Table~\ref{molecular}), 
it is possible to put constraints on the physical state of the gas.
The excitation temperature $T_{0J}$ between rotational levels 0 and J is defined as

\begin{equation}
\frac{N(\rm J)}{N(0)} = \frac{g(\rm J)}{g(0)} e^{-E(\rm 0J)/kT_{0J}}\noindent
\label{excitation}
\end{equation}

where $g(\rm J)$ is the statistical weight of the rotational level J: $g(\rm J) = (2J+1)(2I+1)$ with nuclear spin I~=~0 for even J (para-H$_2$) and I~=~1 for odd J (ortho-H$_2$), $k$ is the Boltzmann constant, and $E(\rm 0J)$ is the energy difference between level J and the ground state (J=0).
If the excitation processes are dominated by collisions, then the populations of the rotational levels follow a Boltzmann distribution described by a unique excitation temperature for all rotational levels. This is generally the case for low rotational levels which have a de-excitation time scale larger than the collision time-scale. Indeed, $T_{01}$ is a good indicator of the kinetic temperature (Roy et al. 2006; Le Petit et al. 2006) especially in clouds similar to the one we study here (log~$N$(H$_2$)~$>$~18). 
However, because of the small energy difference between the J=0 and J=1 levels, the value of $T_{01}$ is very sensitive to uncertainties on $N$(H$_2$,J=0) and $N$(H$_2$,J=1) and the use of higher rotational levels may help derive a better constraint on $T_{\rm K}$. From Fig.~\ref{excitation_H2}, it can be seen that the population of J=0 to J=2 levels can be described by a unique excitation temperature $T_{\rm ex}=69_{-8}^{+10}$~K and $T_{\rm ex}=79_{-10}^{+14}$~K for the first and second component, respectively.

These temperatures are slightly smaller than what was found in previous studies of H$_2$-bearing DLAs (e.g. Ledoux et al. 2003, $T\sim$ 90 to 180~K; Srianand et al. 2005, $T\sim$~153$\pm$78~K), 
but similar to what is measured in the ISM of our Galaxy (77$\pm$17~K; Rachford et al. 2002) and in the Magellanic Clouds (82$\pm$21~K; Tumlinson et al. 2002), where H\,{\sc i} column densities are also large. Note that, temperatures observed through high latitude Galactic sight lines are also larger (124$\pm$8~K; Gillmon et al. 2006, or ranging from 81~K at log~$N$(H$_2$)~=~20 to 219~K at log~$N$(H$_2$)~=~14; Wakker 2006).

The population of J=3 rotational level is in turn enhanced compared to the Boltzmann distribution, which indicates the presence of additional excitation processes such as UV pumping (e.g. Noterdaeme et al. 2007a) and/or turbulent dissipation (as possibly indicated by the larger $b$-parameters for higher-J levels seen by Noterdaeme et al. 2007b).

\subsection{Density}

Absorption lines produced by neutral carbon are seen only in one component at $z_{\rm abs}$~=~3.28746 (see Figure~\ref{ajuste}) and are associated with the strongest H$_2$ component.

We used the relative populations of the two first sub-levels of the C\,{\sc i} ground state to derive the excitation temperature of the C\,{\sc i} fine-structure level, according to the Boltzmann equation
(see Eq.~1).
We have adopted the energy difference between the C\,{\sc i} excited (C\,{\sc i}$^{\star}$: 2s$^2$2p$^2$\,$^3$P1) and true ground-state (2s$^2$2p$^2$\,$^3$P0) levels, $\Delta E_{\rm eg}$~=~23.6~K. The population ratio $N$(J=1)/$N$(J=0) of the C\,{\sc i} fine-structure level corresponds to an excitation temperature of $T_{\rm ex}$ = 15.4$\pm$0.1 K. This is higher than the temperature expected in the case the excitation is dominated by the cosmic microwave background radiation ($T_{\rm CMBR}$~=~11.7~K at $z=3.287$) indicating that excitation by collisions is important. Using the results shown in fig.~12 of Srianand et al. (2005), we derive that the particle density, $n_{\rm H}$, is in the range 50-80~cm$^{-3}$.

\section{Discussion}

We have presented a detailed analysis of a QSO-DLA system with an extremely large column density, 
$\log N($H\,{\sc i}$)=22.0\pm 0.10$, at $z_{\rm abs}=3.287$ towards the quasar SDSS J\,081634$+$144612. The velocity structure of associated metal absorption lines indicates the presence of 8 components grouped into three sub-systems centered at $z=3.28661$, 3.28735 and 3.28814 respectively, spanning $\sim 115$ km~s$^{-1}$. C\,{\sc i} is detected in the $z=3.28735$ sub-system whilst H$_2$ is detected in both the $z=3.28661$ and 3.28735 sub-systems.
From the H$_2$ excitation, we derive a kinetic temperature of $T_{\rm K}\sim 75$ K and the observed column density ratio $N($C\,{\sc i}$)^{\star}/N($C\,{\sc i}$)$ yields a particle density in the range $n_{\rm H}\sim 50-80$~cm$^{-3}$.
The depletion of metals onto dust grains measured in the strongest H$_2$ component located at $z_{\rm abs}=3.28735$ is similar to what is observed in the disc of the Galaxy. All this shows that this system, apart from having unusually large N(H\,{\sc i}), has properties consistent with that of a typical H$_2$-bearing DLA (see Ledoux et al. 2003; Noterdaeme et al. 2008).

While log~$N($H\,{\sc i}$)\geq 22$ DLAs are very rarely seen in front of QSOs, several have already been detected in the optical afterglow spectrum and at the redshift of long-duration $\gamma$-ray bursts. Two of them (towards GRB\,050401, GRB\,080607 and GRB\,060926) even have log~$N($H\,{\sc i}$)>22.5$ (Watson et al. 2006; Prochaska et al. 2009; Jakobsson et al. 2006). It is not surprising to observe a strong DLA at the redshift of GRBs since these objects are expected to be associated with star-forming regions where the gas is likely to be found in large quantities.
Although QSO-DLAs are located close to regions where stars form as indicated by the presence of metals, the detection of C\,{\sc ii}$^{\star}$ absorption and galaxy-like kinematics (Wolfe et al. 2003), their exact nature is not completely elucidated. Some should be associated with the ISM of galaxies especially when molecules are detected (Noterdaeme et al. 2008), others are probably located in the outskirts of galactic haloes (M{\o}ller et al. 2002; M{\o}ller et al. 2004; Fox et al. 2007; Pontzen et al. 2008, Rauch et al. 2008; Rahmani et al. 2010; Fynbo et al. 2010, 2011).

The difference between the populations of GRB-DLAs and QSO-DLAs is apparent because (i) the mean H\,{\sc i} column density is higher in GRBs than in DLAs reaching easily well beyond $10^{21.5}$~cm$^{-2}$ in the former case (see, e.g., Jakobsson et al. 2006; Fynbo et al. 2009) whilst QSO-DLA with log~$N($H\,{\sc i}$)\ge 22$ are very rare (Noterdaeme et al. 2009b), and (ii) the mean metallicity is larger for GRB-DLAs (Savaglio et al. 2006; Fynbo et al. 2006; Prochaska et al. 2007; Fynbo et al. 2008), about 
0.1~solar at $z>2$ (to be compared to $\sim 0.03$~solar for intervening DLAs). Note that molecules (H$_2$, CO) have been detected in a number of QSO-DLAs (Noterdaeme et al. 2008, 2011, Srianand et al. 2008), whereas they are rarely seen in GRB-DLAs (Ledoux et al. 2009). This may be a consequence of small number statistics however (see Fynbo et al. 2006 and Prochaska et al. 2009) and possibly of inadequate data as high spectral resolution in the blue is usually needed (see Ledoux et al. 2009).

\begin{figure}
\centering
\includegraphics[scale=1.1]{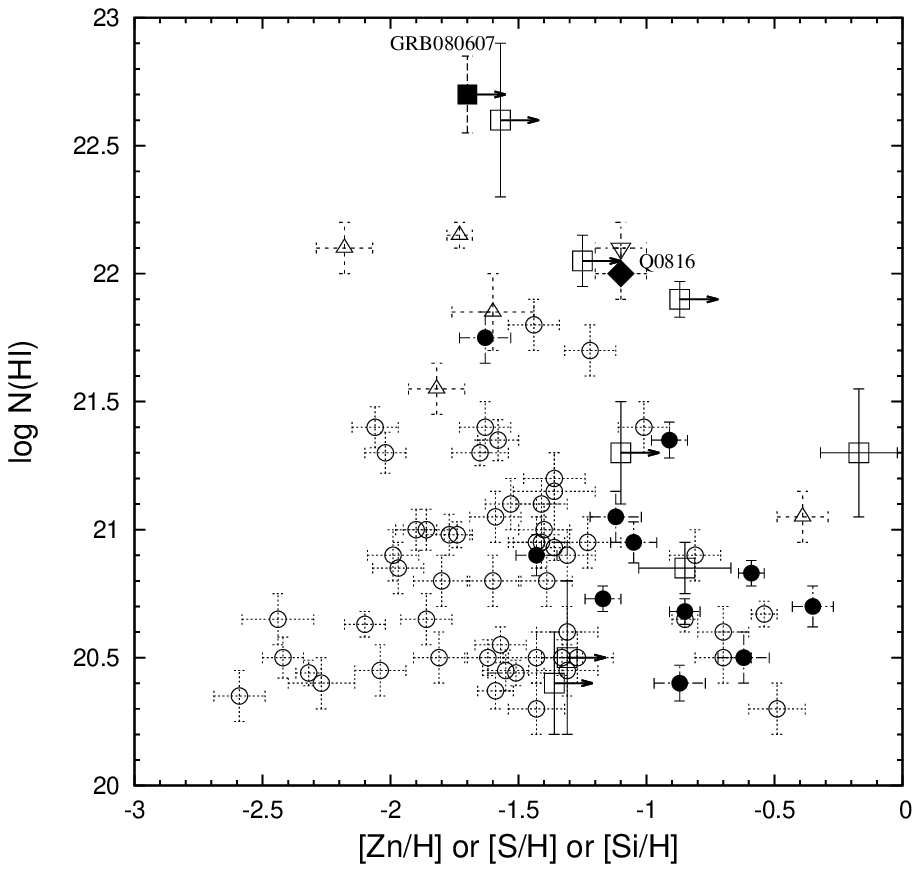} 
\caption{Logarithm of the total neutral hydrogen column density versus metallicity. Circles correspond to QSO-DLAs (Noterdaeme et al. 2008), squares to GRB-DLAs (Prochaska et al. 2009), triangles to GRB-DLAs(Ledoux et al. 2009), inverted triangle to the QSO-DLA towards SDSS J\,113520$+$001053 (Noterdame et al. 2012) and the diamond indicates our measurement for the QSO-DLA towards SDSS J\,081634$+$144612. Filled symbols indicate systems in which H$_2$ is detected, the filled square corresponds to the GRB080607 (Prochaska et al. 2009).}
\label{NHI_MH}
\end{figure}

\begin{figure}
\centering
\includegraphics[scale=1.1]{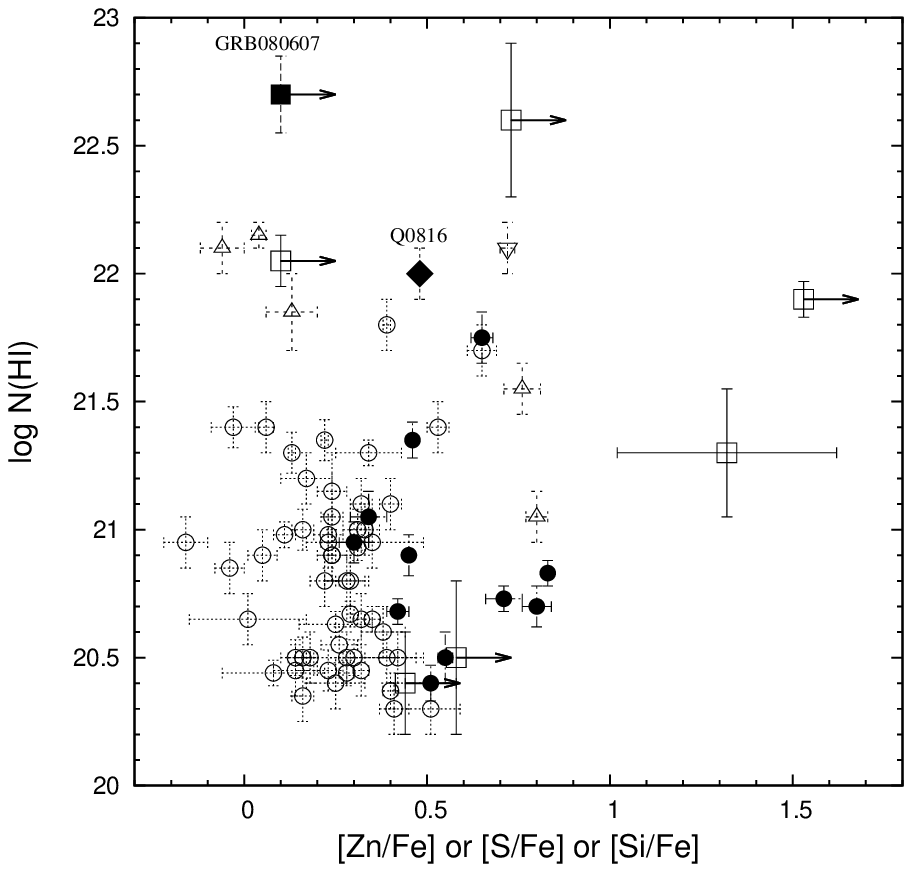} 
\caption{Logarithm of the total neutral hydrogen column density versus depletion. 
Circles correspond to QSO-DLAs (Noterdaeme et al. 2008), squares to GRB-DLAs (Prochaska et al. 2009), triangles to GRB-DLAs (Ledoux et al. 2009), inverted triangle to the QSO-DLA towards SDSS J\,113520$+$001053 (Noterdame et al. 2012) and the diamond indicates our measurement for the QSO-DLA towards SDSS J\,081634$+$144612. Filled symbols indicate systems in which H$_2$ is detected, the filled square corresponds to the GRB080607 (Prochaska et al. 2009) .}
\label{depletion}
\end{figure}

If these differences exist when comparing the overall populations, they may not be that apparent if  we restrict ourselves to high H\,{\sc i} column density systems.
This is why it is interesting to compare the properties of SDSS~J\,081634$+$144612 with those of GRB-DLAs. In Fig.~\ref{NHI_MH}, we plot the logarithm of the H\,{\sc i} column density versus metallicity for QSO-DLAs where H$_2$ is not detected (open circles), QSO-DLAs where H$_2$ is detected (filled circles), the same for GRB-DLAs (squares) and SDSS J\,081634$+$144612 (filled diamond). 
It can be seen that there is a lack of systems with both a high metallicity and a high column density. This is well known for QSO-DLAs and could be a consequence of these DLAs being missed because of the high induced attenuation which makes the QSO drop out of the sample (e.g. Boiss\'e et al. 1998). 
Altough GRB afterglows are for a little while much brighter than quasars, the presence of a dust-bias could also affect their statistics (see e.g. Fynbo et al. 2009; Ledoux et al. 2009; and Greiner et al. 2011). 
A possibility is that GRB-DLA metallicities could be higher than measured as often only lower limits are derived from intermediate resolution observations (see Prochaska 2006, Petitjean \& Vergani, 2011).
This can however probably not explain this lack of systems completely. SDSS J\,081634$+$144612 is located at the limit of the region where systems are missing and within the region where GRB-DLAs are located. 
In Fig.~\ref{depletion}, the logarithm of the H\,{\sc i} column density is plotted for the same systems versus the depletion of iron onto dust grains. 
We have scaled the depletion of chromium in SDSS J\,081634$+$144612. It can be seen here again that the DLA towards SDSS J\,081634$+$144612 is well within the region where GRB-DLAs are located. Therefore, our study indicates that extremely large H\,{\sc i} column density DLAs found toward QSOs have properties similar to those of GRB-DLAs.
\par\noindent
 Following Schaye (2001), Krumholz et al (2009) proposed a radiation-insensitive model --hence 
applying to both QSO and GRB-DLAs-- where the lack of high-$N$(H\,{\sc i}), high-metallicity is explained by the conversion from atomic to molecular gas. Here, the physical conditions in the $H_2$-bearing cloud indicate that we are still observing diffuse gas. This is not in tension with the above model since dense, cold and molecular gas is expected to have a small cross section and is not easily intercepted by the line-of-sight. Interestingly, a fully molecular cloud has been observed in the case of the $\log N$(H\,{\sc i})~=~22.7 DLA associated to GRB080607  (Prochaska et al. 2009), which could be observed very quickly ($<$1\,h) after the burst. Therefore, while the absorption properties of
high-$N$(H\,{\sc i}) GRB and QSO-DLAs  appear to be similar, detecting fully molecular clouds will remain challenging in the case of QSOs, because of the random distribution of the lines-of-sight and the high induced extinction.

Very little is known about GRB host galaxies at high redshift ($z>2$, Savaglio et al. 2009). Cosmological simulations show that GRB-DLAs are predominantly associated with haloes of mass 10$^{10}<M_{\rm vir}$/$M_{\odot}<10^{12}$, an order of magnitude larger than the galaxies responsible for the bulk of QSO-DLAs (Pontzen et al. 2010; see however Barnes \& Haehnelt 2010). 
But what is true for the overall QSO-DLA population does not hold for its high column density end. We have shown here that DLAs with the highest H\,{\sc i} column densities, seem to have absorption properties similar to that of GRB-DLAs. It would be most interesting to search for emission lines associated to these DLAs which have probably small impact parameters (see e.g. Noterdaeme et al. 2012)\footnote{The galaxy responsible for the recently found $\log N($H\,{\sc i}$) \sim 22$ DLA towards J\,1135$-$0010 was detected at $b\approx0.1$~arcsec from the background QSO (Noterdaeme et al. 2012). In this particular case, Kukarni et al. (2012) demonstrated that detecting the Lyman-$\alpha$ emission is even possible from UVES data. 
Here, the DLA is covered at the edge of two UVES echelle orders and possibly contaminated by scattered light in the red arm, preventing us from putting any meaningful limit on the Ly-$\alpha$ flux --within the slit. Follow-up observations specifically tuned to the search of emission lines is thus desirable (see e.g. Fynbo et al. 2010, 2011, P\'eroux et al. 2012).}

Note that the above conclusion is probably not very surprising as high H\,{\sc i} column densities likely arise in gas located in the inner regions of star forming galaxies of moderate attenuation. This idea should be investigated with larger samples of both QSO and GRB-DLAs with high column densities. The BOSS survey (Eisenstein et al. 2011) will soon increase the number of such QSO-DLAs by an order of magnitude.

\acknowledgments
We thank the anonymous referee for helpful comments and suggestions which improved this paper.
PPJ and RS acknowledge the support of the Indo-French Centre for the Promotion of Advanced Research (Centre Franco-Indien pour la Promotion de la Recherche Avanc\'ee) under contract no. 4304-2. S. L. is supported by FONDECYT grant No. 1100214.

\newpage
\onecolumn

%%%%%%%%%%%

\begin{table}[] 
\begin{center}
\footnotesize \small
\caption{Results of Voigt profile fitting analysis.}
\label{parameters}
\renewcommand{\tabcolsep}{2.0pt}
\begin{tabular}{ c c c c c c c c c c c }
\hline
\hline
{\large \strut}$z$     & $v$ & $b$        & \multicolumn{8}{c}{$\log N$~(cm$^{-2}$)}  \\
        & (km\,s$^{-1}$)    &  (km\,s$^{-1}$)          & C\,{\sc i}   &C\,{\sc i}$^{\star}$&C\,{\sc i}$^{\star\star}$& Si\,{\sc ii} & Cr\,{\sc ii} & Ni\,{\sc ii} & Zn\,{\sc ii} & Fe\,{\sc ii}\\
\hline
3.286481&-68  &8.1$\pm$0.2 &              &              &                       &15.73$\pm$0.03&13.43$\pm$0.02&13.87$\pm$0.01&13.04$\pm$0.02&15.09$\pm$0.04\\
3.286746&-50  &8.2$\pm$0.3 &              &              &                       &15.99$\pm$0.04&13.71$\pm$0.02&14.15$\pm$0.01&13.12$\pm$0.02&15.48$\pm$0.03\\
3.286998&-32  &7.4$\pm$1.7 &              &              &                       &14.90$\pm$0.10&12.68$\pm$0.11&13.18$\pm$0.10&12.19$\pm$0.10&14.79$\pm$0.04\smallskip\\
3.287260&-14  &8.7$\pm$1.0 &              &              &                       &14.99$\pm$0.05&12.83$\pm$0.06&13.44$\pm$0.04&12.40$\pm$0.05&14.85$\pm$0.04\\
3.287456&  0  &7.2$\pm$0.2 &13.43$\pm$0.01&13.24$\pm$0.02&12.47$\pm$0.07         &14.90$\pm$0.04&12.63$\pm$0.08&13.33$\pm$0.03&12.40$\pm$0.07&14.57$\pm$0.06\smallskip\\
3.288098&+45  &20.6$\pm$0.8&              &              &                       &15.29$\pm$0.04&13.23$\pm$0.05&13.73$\pm$0.03&12.45$\pm$0.06&14.91$\pm$0.07\\
3.288169&+50  &4.4$\pm$0.6 &              &              &                       &14.89$\pm$0.07&12.69$\pm$0.09&13.23$\pm$0.05&11.73$\pm$0.16&14.41$\pm$0.13\\
3.288345&+62  &6.1$\pm$3.7 &              &              &                       &14.30$\pm$0.25&              &12.30$\pm$0.48&11.34$\pm$0.38&14.82$\pm$0.05\\
\hline
\multicolumn{3}{c}{Total}  &13.43$\pm$0.01&13.24$\pm$0.02&12.47$\pm$0.07         &16.31$\pm$0.01&14.07$\pm$0.02&14.54$\pm$0.01&13.53$\pm$0.01&15.89$\pm$0.02\\
\hline
\end{tabular}
\end{center}
\end{table}

\begin{table}[] 
\begin{center}
\caption{Molecular hydrogen column densities and excitation temperatures}
\label{molecular}
\begin{tabular}{  c c c c c l }
\hline
\hline
$z_{\rm abs}$ 	 & $v$~$^1$ &  J   &   log $N($H$_2$,J)~$^2$ 	& $b$ 	& $T_{\rm ex}$     \\
		 &(km\,s$^{-1}$)        &	   &	  (cm$^{-2}$)		& (km\,s$^{-1}$)		& (K) 		 \\
\hline
3.28667~$^3$     & 55.0 &   0.0     &   16.59$\pm$0.50         &   $2_{-1}^{+2}$		& \multirow{3}{0pt}{\mbox{$T_{\rm ex}^{012} = 69_{-8}^{+10}$}}	 \\
                 & 	&   1     &   17.55 $\pm$0.30         &   $2_{-1}^{+2}$	        &                      		                         \\
		 &      &   2     &   14.71 $\pm$0.20         &   $8.3_{-0.4}^{+0.4}$	&                                                        \\
		 &      &   3     &   15.13 $\pm$0.10         &   $6.5_{-0.4}^{+0.4}$	& $T_{\rm ex}^{03}=158_{-28}^{+44}$                            \\
\hline
3.28742~$^4$     & 2.5  &   0     &   18.19 $\pm$0.35         &   $2_{-1}^{+2}$		&  \multirow{3}{0pt}{\mbox{$T_{\rm ex}^{012}=79_{-10}^{+14}$}}    \\
    	         &      &   1     &   18.41 $\pm$0.20         &   $2_{-1}^{+2}$		&                      	                             	\\
    	         &      &   2     &   16.21 $\pm$0.25         &   $8.3_{-0.4}^{+0.4}$	&                     	                             	\\
    	         &      &   3     &   15.75 $\pm$0.10         &   $6.5_{-0.4}^{+0.4}$	& $T_{\rm ex}^{03}=117_{-12}^{+16}$      	        	\\
\hline
\end{tabular}
\end{center}
$^1$ With respect to C\,{\sc i}.\\
$^2$ The errors on the H$_2$ column densities correspond to the best fits using the range of $b$-values.\\
$^3$ Total log $N$(H$_2$)= 17.60.\\
$^4$ Total log $N$(H$_2$)= 18.62. 
\end{table}

\end{document}